# Separating mixtures of single-fiber orientation distributions for increased sensitivity to crossing white matter fibers

Adam M. Saunders (adam.m.saunders@vanderbilt.edu),[a] Gaurav Rudravaram (gaurav.rudravaram@vanderbilt.edu),[a] Michael E. Kim (michael.kim@vanderbilt.edu),[b] Elyssa M. McMaster (elyssa.m.mcmaster@vanderbilt.edu),[a] Yifan Shuai (yifan.shuai@vanderbilt.edu),[c] Warren Boschen (warren.w.boschen@vanderbilt.edu),[c] Sophie Lin (sophie.c.lin@vanderbilt.edu),[b] Suzana Herculano-Houzel (suzana.herculano@vanderbilt.edu),[d] Daniel C. Colvin (daniel.colvin@vumc.org),[e] Yurui Gao (yurui.gao@vanderbilt.edu),[c,e] Kurt G. Schilling (kurt.g.schilling.1@vumc.org),[a,e,f] Adam W. Anderson (adam.anderson@vanderbilt.edu),[c,e,f] Bennett Landman (bennett.landman@vanderbilt.edu)[a-c,e,f]

[a]Department of Electrical and Computer Engineering, Vanderbilt University

[b]Department of Computer Science, Vanderbilt University

[c]Department of Biomedical Engineering, Vanderbilt University

[d]Department of Psychology, Vanderbilt University

[e]Vanderbilt University Institute of Imaging Science, Vanderbilt University

[f]Department of Radiology and Radiological Sciences, Vanderbilt University Medical Center

**Running title:** Unmixing white matter fibers

## Abstract

Intro: Diffusion MRI allows for in vivo probing of brain tissue architecture via micron-level diffusion aggregated at the millimeter scale, but crossing white matter fibers decrease the sensitivity of downstream tasks such as microstructural modeling and tractography. Current fixel-based methods for identifying the number, orientation, and spread of fibers rely on finding the peaks of fiber orientation distributions from constrained spherical deconvolution. However, this method is biased by the spreads of nearby fibers and fails at angular separations lower than approximately 40-50 degrees.

Methods: In this work, we perform a constrained nonlinear optimization called fiber segmentation by signal splitting and iterative least squares estimation (FISSILE), modeling each voxel as a mixture of single-fiber distributions. We use FISSILE to fit axially symmetric fibers in simulations and on three in vivo human subjects from the Human Connectome Project, an ex vivo raccoon brain, as well as the same human subject scanned twice at two different signal-to-noise ratios (SNRs).

Results: On simulated data, FISSILE successfully identifies fibers at an angular separation of 40° and a volume fraction of 0.1 at an SNR as low as 30, while MRtrix3's fod2fixel is limited to 60° separation and 0.1 volume fraction at the same SNR. On acquired data, FISSILE identified 85.8% of white matter voxels on average in human brains and 79.4% in the raccoon with at least one fiber crossing, whereas fod2fixel identified 30.1% and 44.5%. The number of white matter fiber crossings was mildly sensitive to different noise levels in the human subject with two acquisitions.

Conclusion: A method that models the crossing fiber phenomenon as an unmixing problem increases sensitivity to individual fibers for more precise characterization of anatomy.

**Acknowledgments**

We would like to acknowledge Kelly Lambert (University of Richmond, Department of Psychology) for collecting the raccoon specimen used in this study. Data were provided in part by the Human Connectome Project, WU-Minn Consortium (Principal Investigators: David Van Essen and Kamil Ugurbil; 1U54MH091657) funded by the 16 NIH Institutes and Centers that support the NIH Blueprint for Neuroscience Research; and by the McDonnell Center for Systems Neuroscience at Washington University. This work was funded by NIH 1R01NS136743-01A1 and NIH 5U01DA055347-03.
We used generative AI to create code segments based on task descriptions, as well as to debug, edit, and autocomplete code. Additionally, generative AI technologies have been employed to assist in structuring sentences and performing grammatical checks. We take accountability for the review of all content generated by AI in this work.

**Code and Data Availability Statement**

Code and simulation data used in this study will be available upon publication in open source on Zenodo at https://doi.org/10.5281/zenodo.20044173 (SHA-1 hash: 70c5d1d8b08e6d9f558ee5468419612d72735336 for fissile_simulations.tar.gz and 1cdba69911802f72e7ece5a7d4f25304fece4d0c for fissile_v1.1.3.sif).

**Ethics Approval and Patient Consent Statement**

The raccoon (*Procyon lotor*) brain hemisphere was obtained from an animal live-trapped in Larimer County, CO, near the National Wildlife Research Center (NWRC) under approval by the Animal Review Board at the USDA NWRC (QA-2492). All human subject imaging was obtained with informed consent under approval from the institutional review board.

**Key Points**

- White matter fiber crossings decrease the sensitivity of microstructural models
- The peaks of fiber orientation distribution functions are biased and merge at low crossing angles
- Modeling fiber crossings as a mixture of orientation distribution functions allows for increased sensitivity to crossings

## 1. Introduction

Diffusion magnetic resonance imaging (MRI) allows for millimeter-level description of tissue microstructure with the micron-level diffusion motion of water molecules. The restriction of water molecule diffusion along white matter fibers in the brain provides an opportunity to probe and model these fibers in vivo. Fiber modeling provides better characterization for understanding aging, neurodegeneration, brain connectivity, and injury.[1] Modeling white matter fibers in the brain consists of two steps: first, fitting a voxel-level microstructural model to diffusion MRI data to infer fiber orientation, and second, generating fiber trajectories or streamlines whose routes we estimate from the fiber orientation information.[2]

An open issue in diffusion MRI is how to model the complex geometry of white matter fibers from the observed signal in the brain. Such complex geometries are not rare: Jeurissen et al. in a previous study found that crossing fibers occur in 60-90% of voxels in the human brain.[3] Crossings of white matter fibers, as well as fanning, bending, and bottlenecking, complicate downstream modeling of white matter microstructure.[4] For example, using a diffusion tensor model to describe the aggregate diffusion within a voxel allows for first-order metrics such as fractional anisotropy (FA) to describe the degree of diffusion anisotropy. However, in locations with crossing fibers, FA is artificially lower and inaccurately represents the diffusion within the voxel (Fig. 1).[5] Due to the inherent radial symmetry in diffusion MRI acquisitions, it is difficult to distinguish some complex geometries such as fanning and bending at low signal-to-noise ratio (SNR).[6] Spatial resolution also impacts microstructural metrics like FA.[7] The fractal nature of white matter fiber bundles means that crossing fibers are prevalent even at higher spatial resolutions,[8] suggesting that improvements in image acquisition may not resolve these modeling problems.

To account for more complex fiber configurations and interpret the acquired diffusion MRI signal, there exist many higher-order microstructural models.[9] Parametric multi-compartment models such as the "Standard Model" of diffusion MRI,[10] (which generalizes models including neurite orientation density and dispersion imaging (NODDI)[11] and ball-and-stick models),[12] incorporate prior knowledge on the configuration and diffusivities of compartments within a voxel. Non-parametric representations such as constrained spherical deconvolution (CSD)[6], fiber orientation estimated using continuous axially symmetric tensors (FORECAST)[13], and spherical needlets[14] provide fiber orientation distribution functions (fiber ODFs or FODs) that represent the spherical distribution of fibers within a voxel. Commonly, tractography algorithms use fiber ODFs to determine fiber orientation. In particular, CSD only requires prior knowledge of a fiber response function that can be estimated from the diffusion data itself,[2] but difficulty in distinguishing fibers is not limited to fiber ODFs fitted using CSD. For example, spherical needlets require high b-value and high SNR data for stabilized estimation.[14]

The established microstructural models mentioned above describe complex, varying microenvironments, but do not model the properties of individual fibers such as fiber density or fiber spread.[15] Individual fiber bundles within a voxel may be referred to as "fixels"[16], which can be represented in a variety of ways. One method uses multi-tensor models,[17] which represent each individual fixel as a tensor to produce fiber-level microstructural metrics such as FA. Alternatively, one can also fit spherical distributions, e.g. the Bingham distribution, to each fixel to provide fiber metrics of direction, spread, and anisotropy.[18] The fixel segmentation algorithm in MRtrix3's fod2fixel models each fiber within a voxel as a vector with associated metrics including fiber density, fiber dispersion, or fiber amplitude calculated from a multi-fixel ODF segmentation.[19] Dhollander et al. introduced fixel-based analysis, a framework for statistical comparison of fixels from MRtrix3's fod2fixel within a common spatial framework.[20] Fixel-based analysis enables tract-level analysis, such as describing changes along tracts (groups of spatially related fibers) in traumatic brain injury.[21] In addition to assigning fixels to tracts, quantitative fiber properties from fixels can represent more specific microstructural changes that occur in neurodegeneration than values derived from non-fiber-based models.[15]

Many current fixel-based models such as fod2fixel and Bingham fitting assume that the peaks of fiber ODFs fit using CSD describe fiber populations. However, the fiber ODFs cannot represent peaks with less than around 50° angular separation due to the truncated spherical harmonic representation. Additionally, there is a bias between the true fiber directions and the peaks of fiber ODFs due to the spread of nearby fibers, even at angles that appear separable (Fig. 2). Spherically deconvolving the individual fibers from the fiber ODF would not solve these issues, as the convolution operation uses a single filter that assumes each fiber has the same properties. Instead, in this work, we model a voxel-level ODF as a mixture of single-fixel ODFs, i.e., a multi-fixel ODF. For unmixing these fixel ODFs and reliably estimating the underlying fiber populations, we formally define and characterize fiber segmentation by signal splitting and iterative least squares estimation (FISSILE). In simulated data, FISSILE can disentangle fibers at lower angular separations, with high correlation to a reference set. We demonstrate its use beyond simulations in four human subjects and an ex vivo raccoon brain hemisphere to estimate the number of crossing fibers within the brain, comparing with Jeurissen et al.'s previous estimates.[3]

## 2. Materials & Methods

### Theory

To separate multi-fixel ODFs, we use a constrained nonlinear optimization framework for FISSILE. A previous work described FISSILE for use on a simplified mathematical model of simulated ODFs.[22] Here, we describe a generalized version of FISSILE with additional fitting costs that support use within in vivo and ex vivo diffusion MRI, as well as data from simulated tractography phantoms. Figure 3 visualizes the costs and fitting process for FISSILE. FISSILE models a multi-fixel ODF $Y$ as a mixture of rotated single-fixel ODFs $y^{(f)}$ for $f = 1, \ldots, N_f$ fibers, where the individual fibers' $y^{(f)}$ are along the polar axis of the $f$th fiber frame (i.e., coordinate system). This rotation allows us to evaluate single-fixel ODFs in a common angular framework. $Y$ and $y^{(f)}$ are represented in a spherical harmonic basis using coefficients of order $l$ and degree $m$. First, FISSILE determines an initial set of starting fiber orientations using a watershed algorithm to find local spherical maxima.[23] We can calculate the rotation $R^{(f)}$ that transforms a fixel ODF from its fiber frame to the laboratory frame defined by the scanner, i.e., the total multi-fixel ODF is $Y = \sum_f^{N_f} R^{(f)} y^{(f)}$ . Then, we find the fibers corresponding to these orientations using a regularized least-squares fit:

$$\left\{\hat{y}^{(f)}\right\}_{f=1}^{N_f} = \arg\min_{\{\hat{y}^{(f)}\}} \left\| \sum_{f=1}^{N_f} R^{(f)} \hat{y}^{(f)} - Y \right\|^2 + \sum_{f=1}^{N_f} \sum_{\ell,m\neq 0} w_{\ell,m} \left| \hat{y}_{\ell,m}^{(f)} \right|^2 . \tag{1}$$

Note that the second term of Eq. (1) is a regularization which penalizes all $m \neq 0$ terms, encouraging fibers that are axially symmetric with $w_{\ell,m} = 1$ (chosen from a parameter sweep on simulated data). We assume that fiber ODFs are largely symmetric about the axis of the fiber and that asymmetric ODFs represent multiple fiber bundle orientations. However, we do not use a hard constraint to allow for some degree of flexibility.

After calculating the ODFs $\left\{\hat{y}^{(f)}\right\}$, we evaluate the cost of this solution:

$$\mathcal{L} = \lambda_{\text{fit}}\mathcal{L}_{\text{fit}} + \lambda_{\text{shape}}\mathcal{L}_{\text{shape}} + \lambda_{\text{volume}}\mathcal{L}_{\text{volume}} + \lambda_{\text{belt}}\mathcal{L}_{\text{belt}}, \tag{2}$$

where the cost terms are $\mathcal{L}_{\text{fit}}$, $\mathcal{L}_{\text{shape}}$, $\mathcal{L}_{\text{volume}}$, and $\mathcal{L}_{\text{belt}}$, and the corresponding weights are $\lambda_{\text{fit}} = 1, \lambda_{\text{shape}} = 1, \lambda_{\text{volume}} = 1$, and $\lambda_{\text{belt}} = 2$, chosen empirically. The fit cost is the sum of squared errors

$$\mathcal{L}_{\text{fit}} = \left\|\hat{Y} - Y\right\|_2. \tag{3}$$

The shape cost ensures a that the value of the single-fixel ODF sampled along a set of polar angles $\{\theta_k\}$ decreases as the polar angle increases:

$$\mathcal{L}_{\text{shape}} = \max_{\text{f}} \sum_k \left[\max(0, \hat{y}^{(f)}(\theta_{k+1}) - \hat{y}^{(f)}(\theta_k)\right]^2, \tag{4}$$

where $\{\theta_k\} = \{0°, 15°, 80°, 90°\}$. These angles were chosen to avoid angles that exhibit Gibbs ringing. The volume fraction cost penalizes negative volume fractions of the single-fixel ODF (represented by the spherical harmonic coefficient $\hat{y}_{0,0}$):

$$\mathcal{L}_{\text{volume}} = \begin{cases} 1, \text{if } \operatorname{sgn}\left(\hat{y}_{0,0}^{(f)}\right) = -1 \text{ for any } f \\ 0, \text{otherwise} \end{cases} . \tag{5}$$

Finally, the belt cost penalizes large $\hat{y}_{2,0}$ coefficients that would produce a bulging "belt" around the middle of the FOD (i.e., in the plane perpendicular to the fiber axis). This belt occurs where the FOD value is negative (see the inset in Fig. 3):

$$\mathcal{L}_{\text{belt}} = \max_{f}\left(\max\left(0, \frac{\hat{y}_{2,0}^{(f)}}{\frac{1}{L}\sum_{\ell}^{L}\hat{y}_{\ell,0}^{(f)}} - c\right)\right), \tag{6}$$

where $c = 3$ is the largest $\hat{y}_{2,0}$ coefficient we wish to allow without penalty. Note that the shape, volume, and belt costs are based on the "worst-case" fiber. We normalize cost by the $\mathcal{L}_2$ norm of the total multi-fixel ODF and take the square root to get the relative error:

$$\epsilon_{\text{rel}} = \sqrt{\frac{\mathcal{L}}{\|\hat{Y}\|_2}}. \tag{7}$$

We use MATLAB's MultiStart optimization algorithm to find the set of fibers that minimizes this cost.[24] After optimization, we subtract the integral around the belt of each single-fixel ODF (from the center to the first maximum) to improve the estimated $\hat{y}_0^0$ (relaxation-weighted volume fraction) of the fixel. If the relative error is above a threshold (here, 0.05, chosen from a parameter sweep on simulated data), we take the most asymmetric fixel (i.e., the fixel with the largest sum of $m \neq 0$ terms) and split it into two. We chose this threshold since lower thresholds fit too many fibers to lobes from Gibbs ringing. We then refit all of the directions of the fibers and iterate until we have a solution below the cost threshold. Code for FISSILE is available online in open source at https://github.com/MASILab/FISSILE.

**Simulations**

True fiber orientations are difficult to obtain in the absence of phantoms or histology to provide another source of information from which to estimate information about fibers.[25] As a reference set, we used Fiberfox[26] to simulate white matter and provide a validation dataset. We simulated two bundles crossing at 90 degrees with 10,000 fibers each. From these fibers, we created a multi-compartment model with an intra-axonal stick compartment ($d_{\text{intra}} = 0.001\ \text{s/mm}^2$), an extra-axonal zeppelin ($d_{||} = 0.001\ \text{s/mm}^2$ and $d_{\perp} = 0.00025\ \text{s/mm}^2$), and an isotropic Gaussian diffusion compartment ($d_{iso} = 0.001\ \text{s/mm}^2$). We simulated diffusion MRI with the same acquisition parameters as the HCP data along with 15 levels of Rician noise to provide SNRs from approximately 30 to infinity. Then, we fit ODFs using the same method as above and ran FISSILE and fod2fixel. We also simulated diffusion MRI and fit ODFs on each bundle independently in order to provide a reference dataset for comparison. To compare the estimated fixel ODFs and the reference ODFs, we calculated the angular correlation coefficient (ACC), which measures the degree of correlation between two spherical functions.[13]

**Data acquisition**

We used data from three human subjects from the Human Connectome Project - Young Adult[27] and an ex vivo raccoon right hemisphere. For testing sensitivity of the algorithm to SNR, we also acquired two acquisitions from the same human subject in vivo at different noise levels. For the Human Connectome Project, we randomly selected diffusion MRI and an associated T1-weighted structural MRI from three subjects.[27] The diffusion MRI for the HCP data was a high angular resolution multi-shell acquisition with 270 total directions at b-values of 1000, 2000, and 3000 s/mm$^2$ along with 18 $b = 0$ s/mm$^2$ volumes. The diffusion data were collected on a 3T scanner at Washington University at 1.25 mm isotropic resolution using single-shot echo-planar imaging (TE = 89 ms, TR = 5.5 s) alongside a structural T1-weighted image at 0.7 mm isotropic resolution.[28] The diffusion data were corrected for distortions due to susceptibility inhomogeneities as well as eddy currents using reverse phase encoding acquisitions.[29] We ran quality assurance using the PreQual pipeline.[30]

The raccoon (*Procyon lotor*) brain hemisphere was obtained from an animal live-trapped in Larimer County, CO, near the National Wildlife Research Center (NWRC) under approval by the Animal Review Board at the USDA NWRC (QA-2492). The animal was maintained in captivity in an outdoor enclosure; the estimated age of the animal at the time of its death was under 5 years. The other brain hemisphere was part of a previous study.[31] The ex vivo raccoon acquisition was a multi-shell acquisition with 150 directions at b-values 3000 and 6000 s/mm$^2$ along with 15 $b = 0$ s/mm$^2$ volumes. The data were collected on a 9.4T scanner at 0.375 mm isotropic resolution using multi-shot echo-planar imaging (TE = 30.5 ms, TR = 0.5 s). We also acquired a structural T2*-weighted image at 0.25 mm isotropic resolution. No reverse phase encoding acquisition was available to perform distortion correction, but the use of a multi-shot acquisition reduced the impact of eddy currents.[32] The raccoon brain sample was originally acquired under approval from the Institutional Animal Care and Use Committee.

MRI data for the SNR experiment came from one human subject scanned at two different noise levels with a 3T scanner under approval from the Vanderbilt Institutional Review Board. The first acquisition was a high SNR, high angular resolution acquisition at a spatial resolution of 2.3 mm × 2.3 mm × 2.5 mm with 5 $b = 0$ s/mm$^2$ volumes and 195 directions at b-values of 1000 s/mm$^2$, 2500 s/mm$^2$, 4000 s/mm$^2$, and 5500 s/mm$^2$ (TE = 73 ms, TR = 4 s). The second acquisition was slightly lower in SNR, with a spatial resolution of 2 mm × 2 mm × 2 mm, 1 $b = 0$ s/mm$^2$ volumes, and 168 directions at b-values of 800 s/mm$^2$, 1600 s/mm$^2$, and 2400 s/mm$^2$ (TE = 64 ms, TR = 4 s). We also acquired a T1-weighted image at 1 mm isotropic resolution. To calculate the SNR, we found the mean signal in the brain in $b = 0$ s/mm$^2$ volumes compared to the mean noise level in the brain estimated from Veerart et al.'s method using principal components of the diffusion signal.[33] We ran distortion correction using synthetic reverse phase encoding acquisitions from Synb0-DisCo[34] and ran quality assurance with the PreQual pipeline.

**Fiber analysis**

Processing for fiber analysis was consistent for all subjects and the raccoon brain. We estimated a single-fiber white matter, gray matter, and cerebrospinal fluid response function from the diffusion MRI using an unsupervised algorithm.[35,36] To generate ODFs, we used multi-shell multi-tissue constrained spherical deconvolution (MSMT-CSD) with spherical harmonics of maximum order 8.[37] We rescaled and normalized all the ODFs to integrate to 1. To produce a white matter mask for the human subjects, we used a parcellation from the FreeSurfer pipeline (version 7.2.0) calculated using the associated T1-weighted image.[38] For masking the raccoon brain, we applied an unsupervised 3-class tissue segmentation on the structural image[39] and registered the white matter label to the diffusion MRI. To remove spurious voxels, we retained only the largest connected component of the mask.

We used two algorithms for fitting single-fixel ODFs to the multi-fixel white matter ODFs: FISSILE and MRtrix3's fod2fixel algorithm.[40] We did not compare to a Bingham fitting because converting the Bingham representation to a spherical harmonic representation would be essentially equivalent to fod2fixel. Additionally, the fixel-based analysis statistical framework relies on fod2fixel.[20] MRtrix3's fod2fixel performs a peak finding on segmentations of a multi-fixel ODF to identify fiber directions. First, fod2fixel samples the ODF on a mesh of 1281 directions and uses its peak to identify fibers within a voxel. It uses a level set of a sphere of decreasing radius to compare vertices to their neighbors to find local maxima and segment the ODF into lobes. Fod2fixel removes any lobes that have a peak amplitude smaller than a threshold of 0.1 in order to reduce false positives, (e.g., from Gibbs ringing due to the truncated spherical harmonic representation). Typically, fixel-based analyses such as fod2fixel do not consider single-fixel ODFs, instead opting to model each fiber within a voxel as a vector associated with a metric, for example, fiber dispersion, amplitude, or apparent fiber density. However, the lobe segmentations can be transformed into a spherical harmonic representation and serve as single-fixel ODFs for comparison with FISSILE.

We ran both FISSILE and fod2fixel on all voxels in the white matter for each of the three HCP subjects and the raccoon brain. Because FISSILE's non-convex optimization is computationally expensive (with the longest times at 20 minutes), we ran each voxel independently in parallel on a high-performance computing cluster with 1,344 cores and 10.3 TB of total RAM. We ran fod2fixel locally on a single workstation with 20 cores and 64 GB of memory. Both FISSILE and fod2fixel run independently on each voxel within images and do not incorporate learned information across voxels or subjects. After rescaling the single-fixel ODFs by the integral of the total multi-fixel ODF, we quantitatively evaluated the single-fixel ODFs by calculating the apparent fiber density (AFD) for each fiber using both FISSILE and fod2fixel. AFD is the integral of the unnormalized single-fixel ODF, i.e., $y_{0,0}$. At high b-values, we can assume that the total volume of the fibers is proportional to the diffusion signal,[19] so AFD provides a metric for voxelwise comparison of fibers within a subject.

## 3. Results

### Simulation

Simulated ODFs from Fiberfox allow us to compare FISSILE and fod2fixel to reference fixel ODFs fit on individual fiber bundles. From the simulations, we observe that both methods fail to identify fibers that occupy a volume fraction less than 0.2 within a voxel across a range of SNRs (Fig. 4). While fod2fixel fails to identify fibers smaller than an angular separation of 60°, FISSILE can identifies fibers at about 40° angular separation. The likely reason is that the peaks of fibers have merged past about 55° angular separation (see Fig. 2), and fod2fixel requires separable peaks to identify fibers. These simulations demonstrate that FISSILE's main difference in performance is due to increased angular sensitivity, and there is not a major difference in performance across volume fraction.

### Acquired data

Figure 5 visualizes the first three fixel ODFs for a region of interest in an example HCP subject and the raccoon. Qualitatively, the primary directions identified by FISSILE and fod2fixel are largely very similar, although the volume fractions of the largest fixel ODF from FISSILE are somewhat reduced in regions with multiple fibers that fod2fixel does not identify. Within the example subject from HCP, FISSILE identified more secondary and tertiary fibers than fod2fixel. For example, near the most posterior part of the corpus callosum near the intersection with the superior longitudinal fasciculus, FISSILE identifies more crossing fibers than fod2fixel. Many of these fibers identified by FISSILE do not contain an obvious secondary peak for fod2fixel (see the center of the corpus callosum in Fig. 5). Results were similar in the raccoon, albeit to a lesser extent. Overall, we observe in both datasets that the increased angular sensitivity from FISSILE presents best in secondary fibers with close angular separations.

Next, we calculated the number of fibers in each voxel. Following the method of Jeurissen et al.[3], we rejected any fibers with a volume fraction less than 0.05. Figure 6 shows an axial slice of the number of fibers in each voxel for the HCP subjects and raccoon brain. FISSILE identifies more fibers than fod2fixel throughout the brain, especially in the deep white matter where many white matter tracks converge. The superficial white matter near the gyri also contains areas with three or more fibers. On the other hand, fod2fixel identifies only one or two fibers for most voxels. As expected, both methods predict the least number of fibers in the middle of the corpus callosum, which does not intersect with other bundles. With FISSILE, 85.8% of white matter voxels in humans (79.4% in the raccoon) have a secondary fiber, while fod2fixel identifies a secondary fiber in 30.1% of white matter voxels (44.5% in the raccoon). Our results demonstrate that the majority of the white matter contains crossing fibers, which is consistent with findings from Jeurissen et al.

Figure 7 shows AFD across each fiber number from FISSILE and fod2fixel for an example subject from the HCP dataset. The AFD was mostly concentrated in the first fiber for fod2fixel, while the values were more spread out across fibers for FISSILE. This finding suggests that the fixel ODFs identified from fod2fixel represent larger fiber populations across fewer bundles. The difference between the two methods was significant for the first three fibers ($p < 0.05$, Mann-Whitney test).

Next, we compared FISSILE and fod2fixel in the same human subject across two noise levels. Figure 8 visualizes the largest fixel ODFs estimated for both acquisitions as well as the total number of fibers estimated in each voxel. The SNR for the first acquisition was approximately 53.8, while the SNR for the second acquisition was slightly lower at 29.8. Both FISSILE and fod2fixel are slightly sensitive to SNR. With FISSILE, the percentage of white matter voxels with fiber crossings is 80.6% from the higher SNR acquisition and 87.3% from the lower SNR acquisition, while the percentage of white matter voxels with crossings from fod2fixel is 25.7% of voxels from the higher SNR acquisition and 33.0% of voxels from the lower SNR acquisition.

## 4. Discussion

Instead of modeling fibers using peaks on ODFs, FISSILE provides increased sensitivity to detect crossing fibers by modeling a mixture of fixel ODFs and iteratively solving the unmixing problem. Jeurissen et al. previously used peak finding on ODFs estimated from CSD on single-shell diffusion data as well as a multi-compartment model to estimate that the number of crossing fibers within the human brain falls within 60-90% of voxels.[3] Our results from the three human subjects agree with the upper bound of their estimates. Additionally, we found similar estimates for the prevalence of crossing fibers from the ex vivo raccoon brain. Jeurissen et al. also concluded that CSD cannot resolve fibers smaller than about 40°,[3] which we observed in simulation as well.

Here, we used fixel ODFs from FISSILE and fod2fixel to identify the number of fibers within a voxel, but fixel ODFs enable other analyses. For example, Schilling et al. used fod2fixel to identify bottlenecks where fibers meet and then separate – mapping which fixels are shared by multiple bundles rather than which voxels share multiple fixels as in this study.[4] FISSILE could power a similar analysis focusing on the prevalence multi-fixel multi-bundle voxels.

All of the fiber models considered here rely on fiber ODFs fit using CSD, which is useful for fitting fiber ODFs because it requires few assumptions and has formulations for both single- and multi-shell diffusion acquisitions. However, it assumes a single fiber response across tissue when deconvolving to find ODFs. The impact of this fiber response is subtle but can have downstream impacts on tractography.[25] Instead, a microstructural model that accounts for multiple fiber bundles and fit this model directly to diffusion MRI would be useful, as current models do not account for small crossing angles when compared to reference sets from histology.[41] Alternative basis representations for ODFs beyond spherical harmonics such as spherical needlets could also provide more separable representations for ODFs, though this method requires acquisitions with high SNR and high b-values.[14]

One limitation of FISSILE is the computational expense required for performing a voxel-wise non-convex optimization. Given this extensive fitting time, we only validated on three subjects from a widely-used dataset of young, healthy subjects. One of our previous works attempted to accelerate this fitting by estimating a simplified version of FISSILE using a spherical convolutional neural network, but this model was only validated in simulation and still relied largely on peak finding.[22] As an alternative, we could also focus on areas suspected to have crossing fibers instead of fitting all white matter voxels as we did here for completeness. We also chose the error threshold empirically, but a dynamic threshold that adapts based on prior knowledge and/or SNR would be useful. FISSILE is also limited in its ability to distinguish between crossing and fanning fibers, since it penalizes asymmetry within a fixel ODF. At low SNR, certain fiber configurations such as crossings and fannings are difficult to distinguish when considering an independent ODF from a single voxel, though incorporating knowledge from adjacent voxels or bundles could help as done by Schilling et al.[4] Here, we found that the number of fiber crossings estimated with FISSILE was largely insensitive to SNR.

## 5. Conclusion

Complex fiber geometries such as crossing fibers require us to consider sub-voxel microstructural models given voxel-level diffusion signals. Commonly used first-order metrics like FA are useful, but they limit our ability to probe sub-voxel diffusion. Fiber-level models based on fixels turn this limitation into an opportunity for fiber bundle-specific metrics, but previous works have not considered representing fibers as a mixture model. Here, by introducing this representation, FISSILE addresses a limitation in fixel-based analysis, achieving a large increase in the number of crossing fibers detected and about a 50% increase in angular sensitivity. By shifting from peak detection to fiber unmixing, FISSILE increases sensitivity to detect fibers with smaller angles of separation, enabling more precise fiber-based analyses of microstructural anatomy in the brain and uncovering what happens between the peaks of fiber distributions.

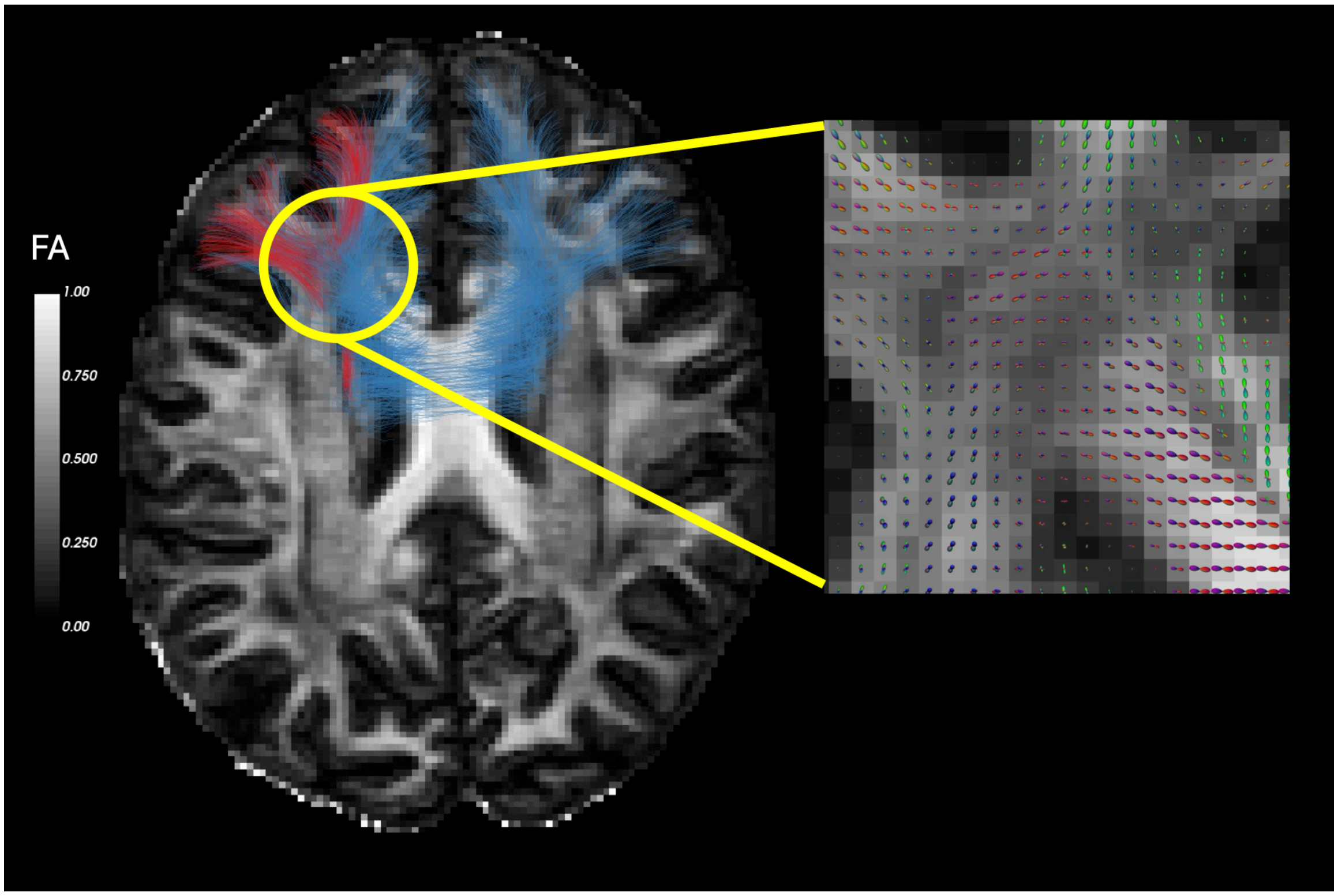

Figure 1: Crossing white matter fibers decrease the sensitivity of microstructural models. For example, crossing fibers at the intersection of the anterior thalamic radiation (red) and the genu of the corpus callosum (blue) artificially decrease fractional anisotropy (visualizations from OpenDIVE[42]).

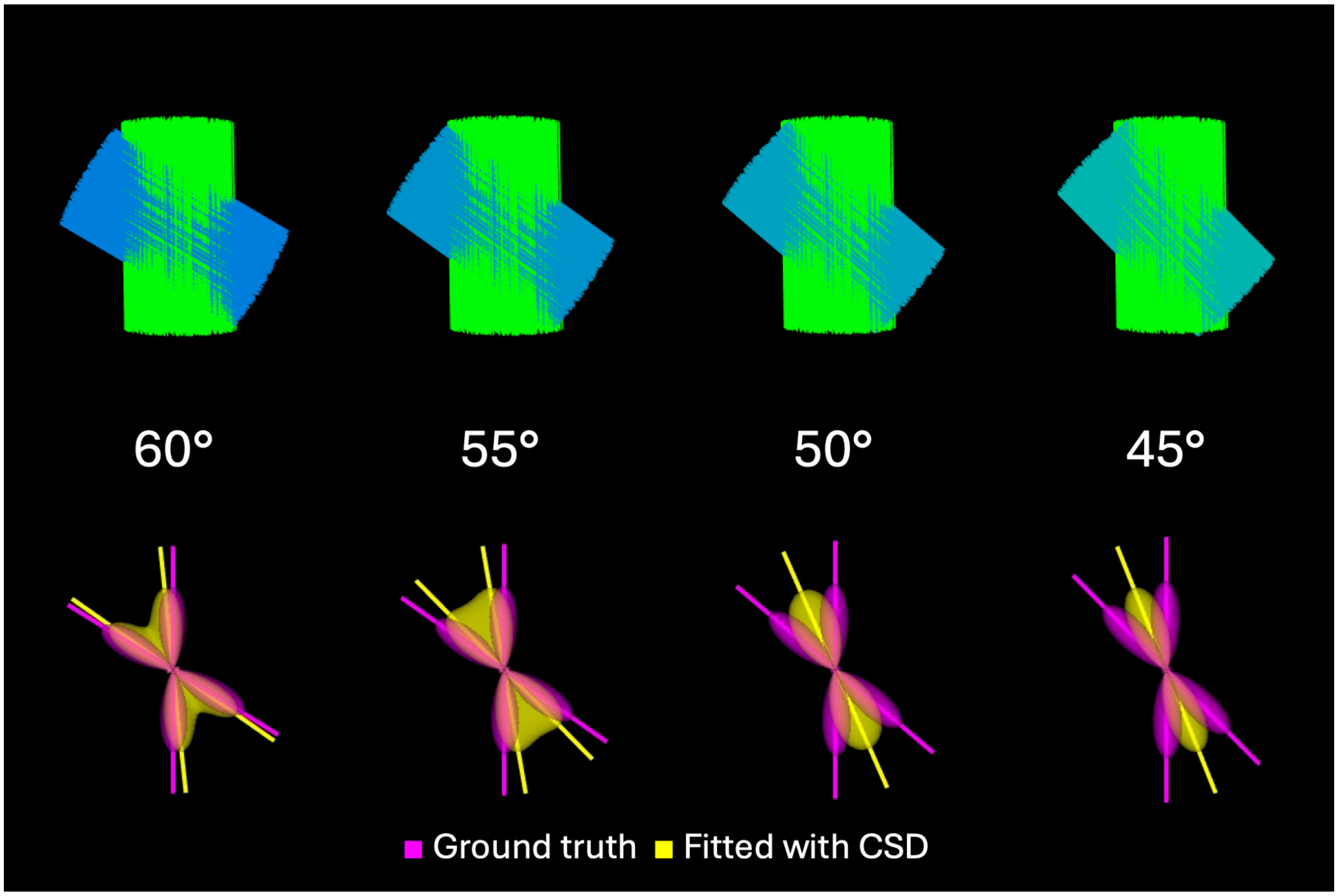

Figure 2: The peaks of fiber ODFs fail to distinguish fiber bundles less than about 50° apart. Even at angles that are separable, the peaks of ODFs are biased relative to the true direction of fiber bundles due to the spread of nearby fibers. (ODFs are fit using diffusion MRI simulated from tractograms using Fiberfox[26] (top row), with ground truth single-fixel ODFs in magenta and multi-fixel ODFs in yellow.)

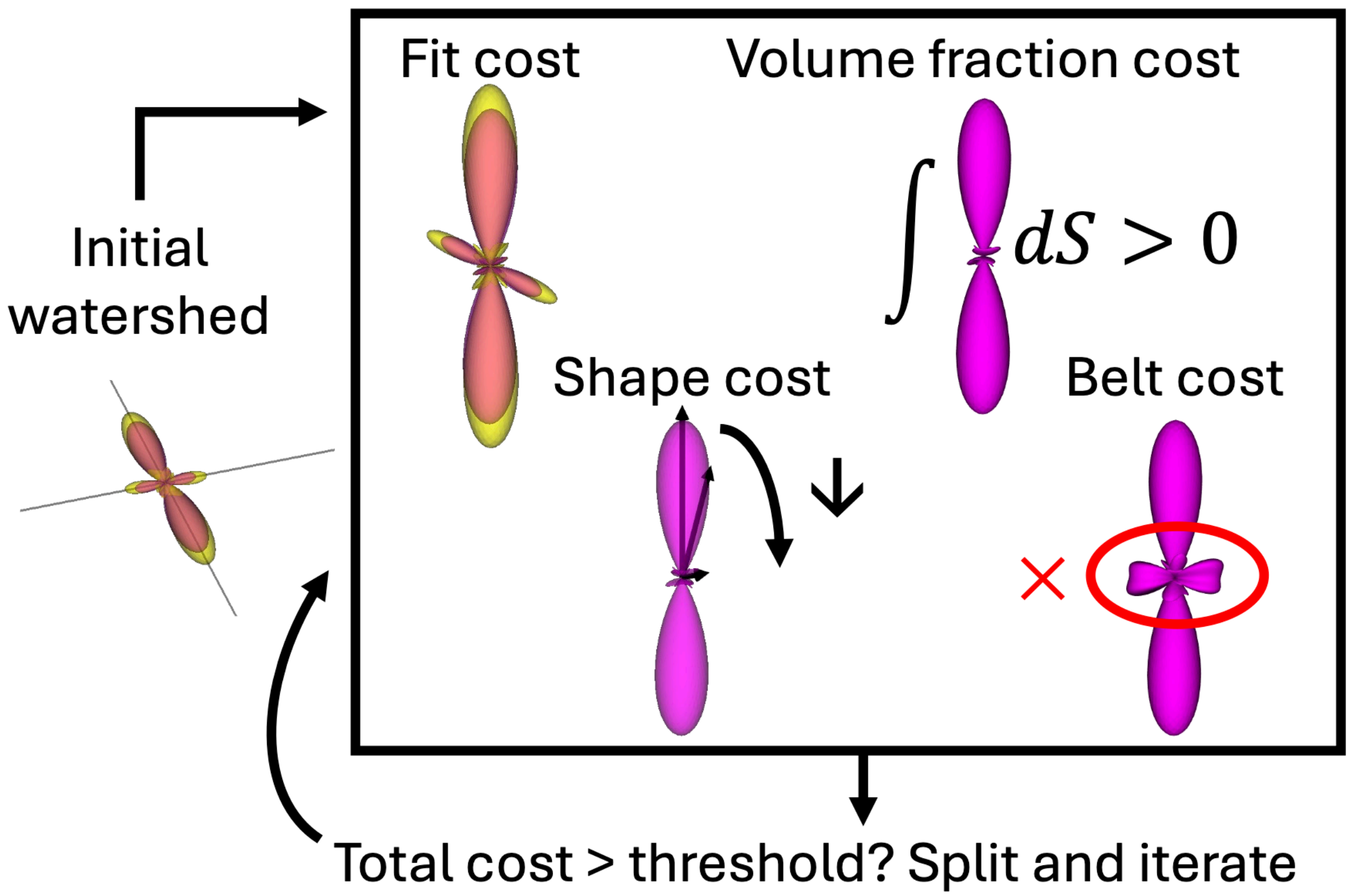

Figure 3: We fit single-fiber (fixel) ODFs around peak fiber orientations from an initial watershed. We evaluate the multi-fixel ODF with costs that encourage goodness of fit to the original ODF and figures of merit for each fixel's ODF: decreasing amplitude as polar angle increases, positive volume fraction, and low amplitude perpendicular to the fixel axis (the "belt" region). If the cost is beyond a threshold, we split the most asymmetric fixel ODF and iterate.

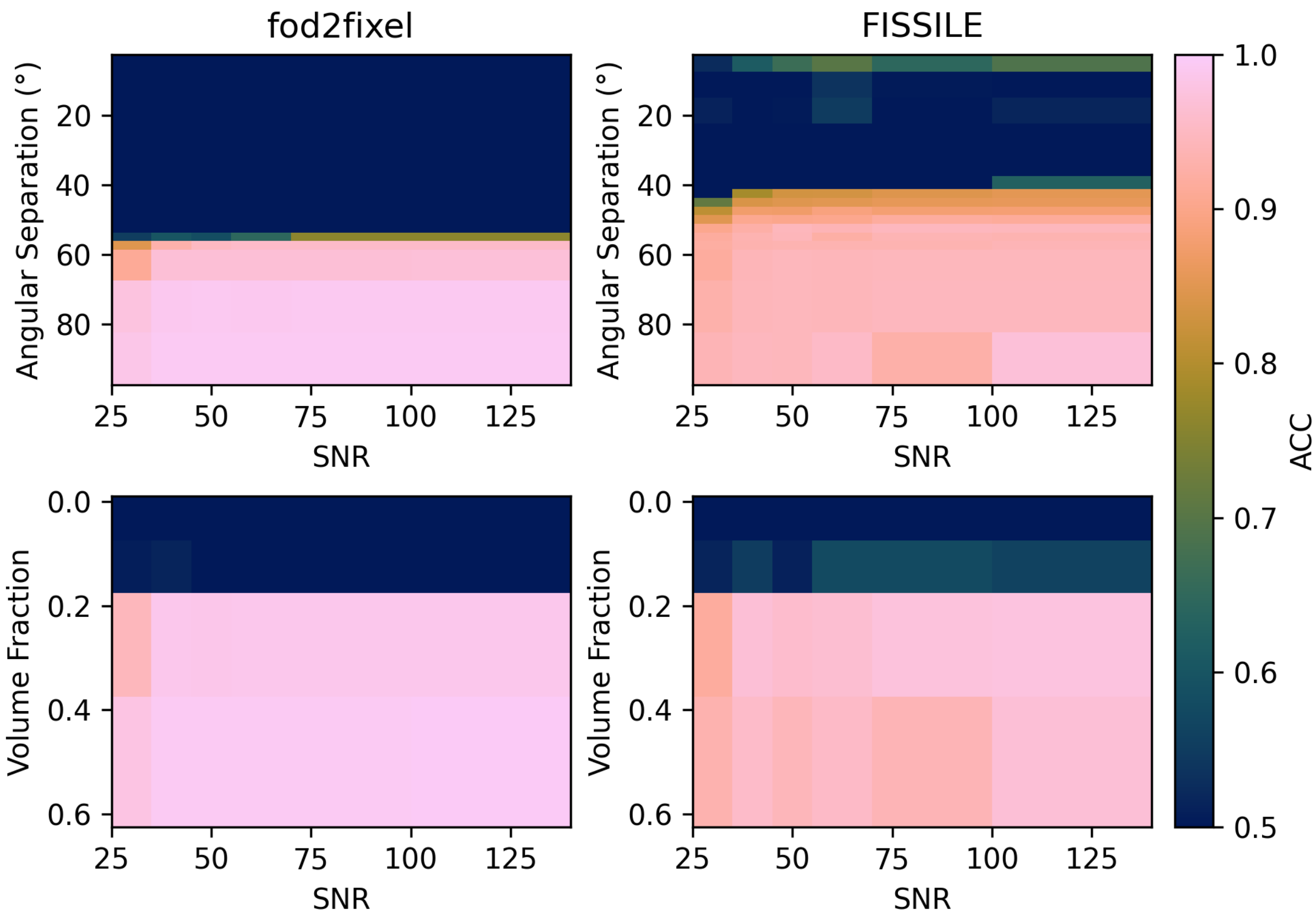

Figure 4: In simulation generated from Fiberfox, the angular correlation coefficient (ACC) of identified fixel ODFs drops below 0.9 at about 60° angular separation and 0.2 volume fraction using fod2fixel at an SNR of 30, but FISSILE distinguishes fibers below 40° angular separation and 0.2 volume fraction.

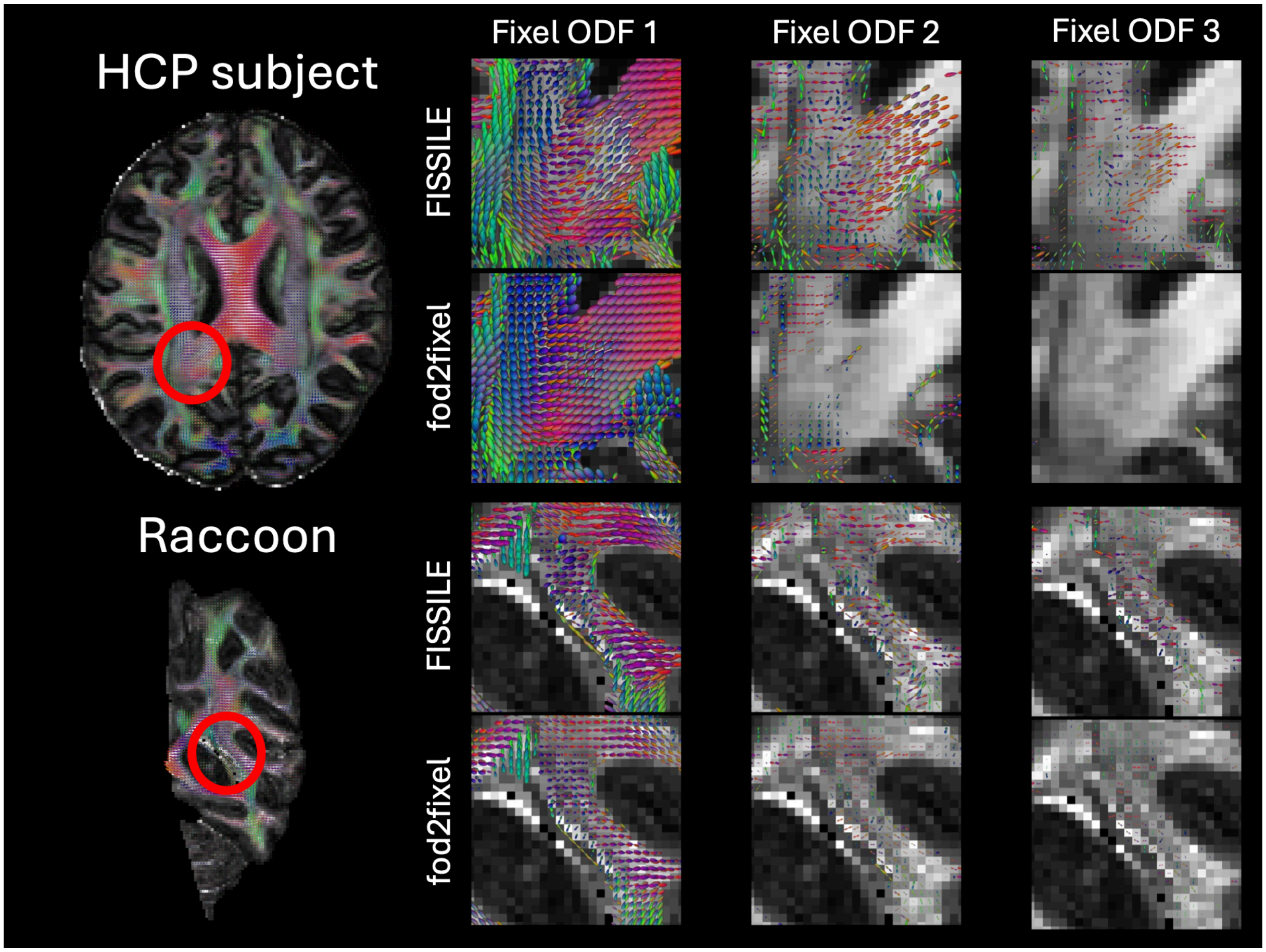

Figure 5: FISSILE identifies secondary fibers in voxels with no secondary peak (i.e., unresolved fiber crossings using fod2fixel); for example, at the crossing of the corpus callosum and superior longitudinal fasciculus in an example subject from the HCP dataset.

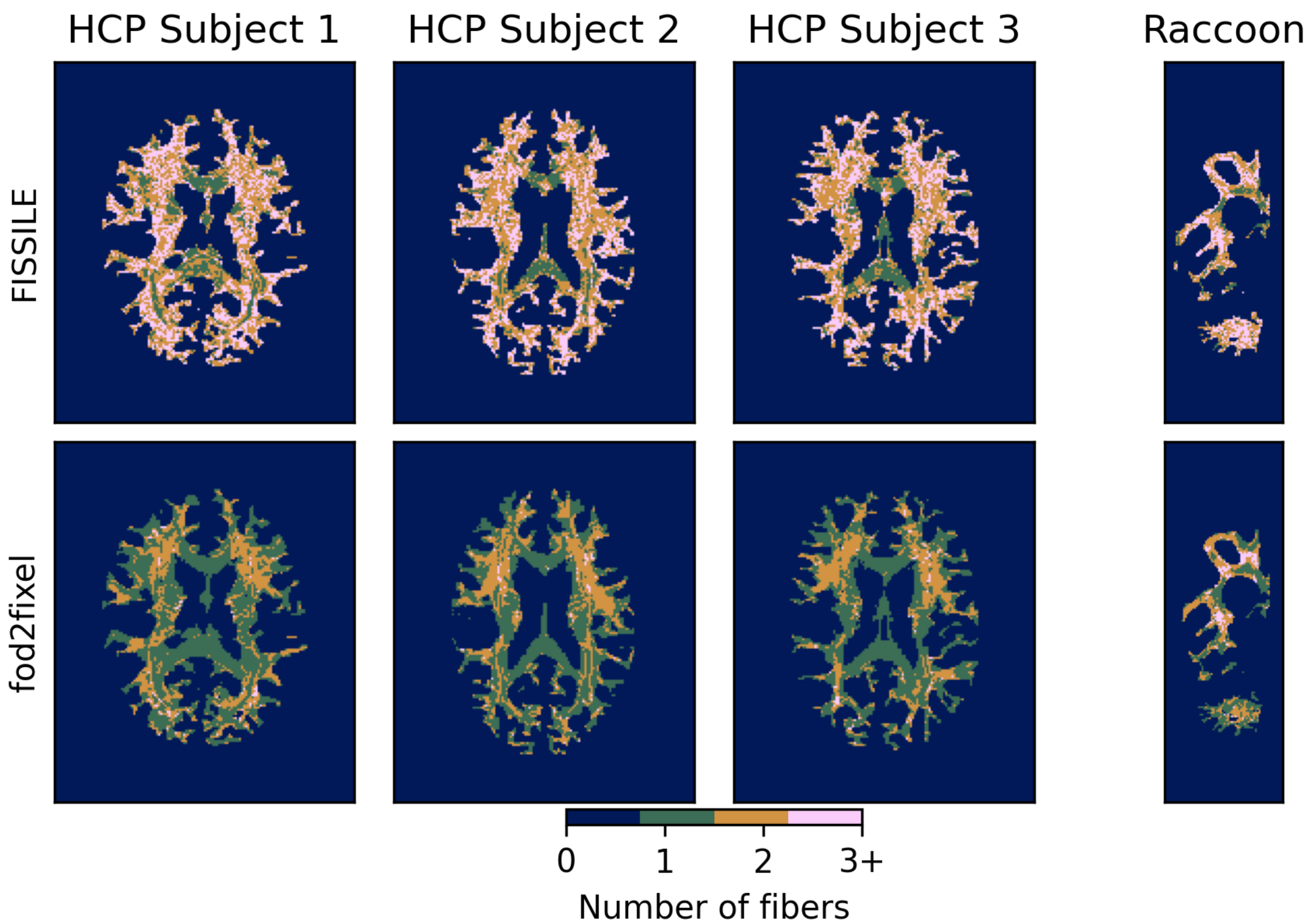

Figure 6: Compared to fod2fixel, FISSILE predicts more fibers throughout the brain in 3 example HCP subjects and an ex vivo hemisphere from a raccoon, especially in the deep white matter. Using FISSILE, we find that 85.8% (79.4% for raccoon) of voxels have at least two fibers, while fod2fixel estimates 30.1% (44.5% for raccoon).

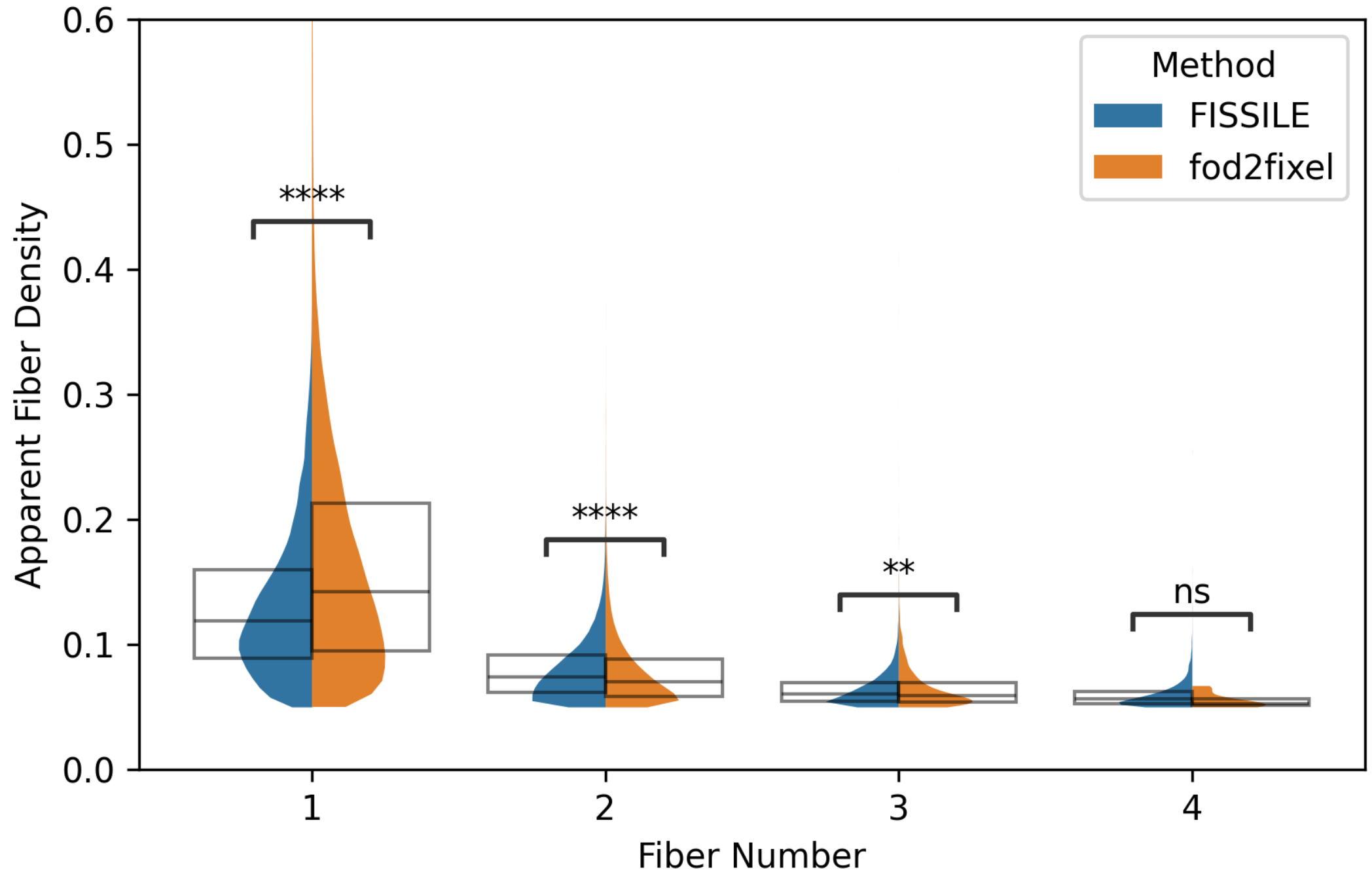

Figure 7: Apparent fiber density (i.e., the integral of a fixel ODF) calculated from FISSILE is significantly different than fod2fixel across the largest three fibers (****: $p < 0.0001$, **: $p < 0.01$, Mann-Whitney test). FISSILE identifies smaller fibers, so the fiber density is more spread out across the fibers.

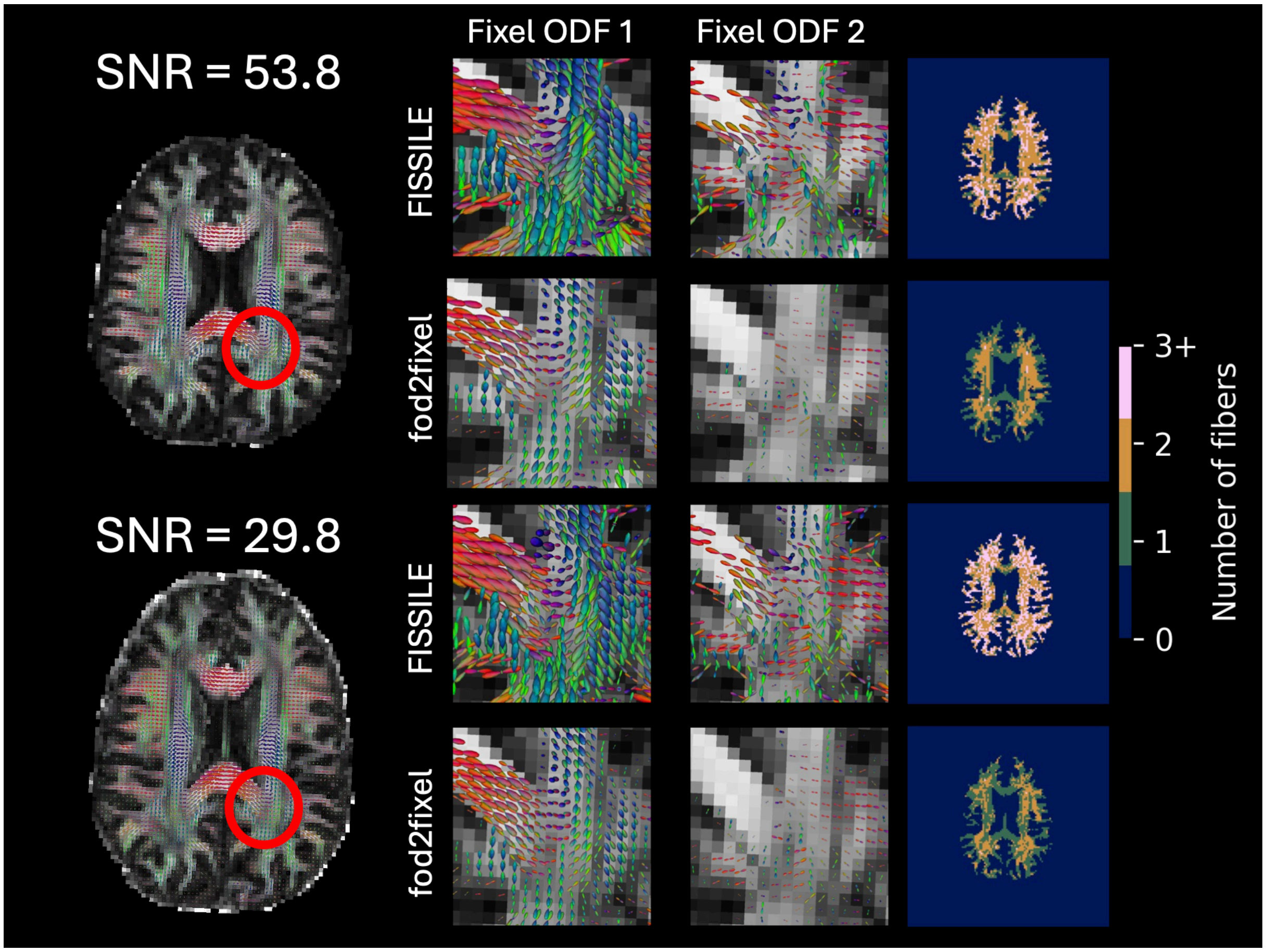

Figure 8: In the same human subject across two acquisitions at different noise levels, the estimated fixel ODFs are similar in both FISSILE and fod2fixel, though the total number of fibers in each voxel is slightly different depending on SNR (80.6% from the higher SNR acquisition and 87.3% from the lower SNR, while fod2fixel estimates 25.7% and 33.0%, respectively).